\documentclass[letter]{aa} 

\usepackage{graphicx}
\usepackage{txfonts}
\usepackage[]{hyperref}
\usepackage{xcolor}

\begin{document}

   \title{Magnetic-buoyancy-induced mixing in AGB stars: a theoretical explanation of the non-universal relation of [Y/Mg] to age }\titlerunning{Mixing in AGB stars induced by magnetic buoyancy}
   \authorrunning{Magrini et al.}

   \author{Magrini L.\inst{1}, Vescovi D.\inst{2,3}, Casali G.\inst{1,4}, Cristallo S.\inst{5,3},  Viscasillas V\'azquez C.\inst{6},  Cescutti G.\inst{7,8}, Spina L.\inst{9}, Van Der Swaelmen M.\inst{1}, Randich S.\inst{1}
          }

\institute{ INAF - Osservatorio Astrofisico di Arcetri, Largo E. Fermi 5, 50125, Firenze, Italy \email{laura.magrini@inaf.it}
\and 
Goethe University Frankfurt, Max-von-Laue-Strasse 1, Frankfurt am Main 60438, Germany
\and 
INFN - Sezione di Perugia, Via A. Pascoli snc, 06123, Perugia, Italy
\and
Dipartimento di Fisica e Astronomia, Universit\`a degli Studi di Firenze, via G. Sansone 1, 50019 Sesto Fiorentino (FI), Italy
\and
INAF - Osservatorio Astronomico d'Abruzzo, Via Mentore Maggini snc, 64100, Teramo, Italy
\and
Institute of Theoretical Physics and Astronomy,
Vilnius University, Sauletekio av. 3, 10257 Vilnius, Lithuania
\and
INAF - Osservatorio Astronomico di Trieste, Via G.B. Tiepolo 11, 34143, Trieste, Italy
\and
IFPU, Institute for the Fundamental Physics of the Universe, Via Beirut 2, 34151, Grignano, Trieste, Italy
\and
INAF - Osservatorio Astronomico di Padova, vicolo dell'Osservatorio 5, 35122, Padova, Italy}

   \date{Received; Accepted}

 
  \abstract
   {Abundance ratios involving Y or other slow-neutron capture elements are routinely used to infer stellar ages.}
   {We aim to explain the observed [Y/H] and [Y/Mg] abundance ratios of star clusters located in the inner disc with a new prescription for mixing in asymptotic giant branch (AGB) stars. }
   {In a Galactic chemical evolution model, we adopted a new set of AGB stellar yields in which magnetic mixing was included. We compared the results of the model with a sample of abundances and ages of open clusters located at different Galactocentric distances.}
   {The magnetic mixing causes a less efficient production of Y at high metallicity. A non-negligible fraction of stars with super-solar metallicity is produced in the inner disc, and their Y abundances are affected by the reduced yields. The results of the new AGB model qualitatively reproduce the observed trends for both [Y/H] and [Y/Mg] versus age at different Galactocetric distances.  }
   {Our results  confirm from a theoretical point of view that the relation between [Y/Mg] and stellar age cannot be 'universal', that is, cannot be the same in every part of the Galaxy. It has a strong dependence on the star formation rate, on the s-process yields, and on their relation with metallicity, and it therefore varies throughout the Galactic disc.   }


   \maketitle
%

\section{Introduction}

The launch of the {\em Gaia} satellite, with its first data releases \citep{gaiadr1, gaiadr2a, gaiadr2b, gaiadr3}, together with ground-based large spectroscopic surveys such as the Apache Point Observatory Galactic Evolution Experiment (APOGEE) \citep{Majewski17}, the Gaia-ESO survey \citep{Gil, randich13}, the GALactic Archaeology with HERMES (GALAH) survey  \citep{gala,galah2018}, and the Large Sky Area Multi-Object Fibre Spectroscopic Telescope (LAMOST) survey \citep{lamost,lamost2} have inaugurated a new era for Galactic astronomy. In the new multidimensional view of our Galaxy, the determination of the stellar ages plays a predominant role. 
New techniques based on the relations between the age and some stellar properties, for instance, 
asteroseismology, gyrochronology, stellar activity \citep[see][for reviews on the issue]{Soderblom10,soderblom14}, and abundance ratios \citep[e.g.][]{masseron15, feltzing17,spina18,casali19,delgado19} have been developed as alternative and complementary methods to the isochrone fitting. 
In this framework, the measurements of stellar ages through the  so-called chemical clocks, that is, abundance ratios composed of a pair of elements that display an increasing trend with decreasing stellar ages, such as [Y/Mg], play a key role. 
Among the first works focusing on these kinds of ratios, \citet{dasilva12} found a significant trend of [Y/Mg], [Sr/Mg], [Y/Zn], and [Sr/Zn] with the age of a sample of solar-type stars. \citet{nissen15} attributed the trend to the different production mechanisms of the pair of elements considered in the ratios: Mg, Zn, and for example, Al and Ti \citep[used  in some subsequent works, as][]{delgado19, casali20} are also produced by massive stars on shorter timescales, while Y, Sr,  and the other s-process elements are synthesised in low- and intermediate-mass stars and are released to the interstellar medium at later times.
The value of their ratios therefore increases in the younger stellar populations, and it has been used as a sort of  chemical clock to infer stellar ages, mainly in the solar neighbourhood \citep[see e.g.][]{jofre20, nissen20}.

 \citet[][hereafter C20]{casali20} analysed a sample of solar-type stars, spanning a wide range of metallicities, to derive precise stellar parameters and abundances  from their high-resolution
stellar spectra. From these, stellar ages  were derived through isochrone fitting. C20 obtained a set of relations between ages and several abundance ratios taking the metallicity dependence into account. 
The relations  were applied to a sample of open clusters observed by the Gaia-ESO survey \citep{Gil, randich13} that were located in a wide range of Galactocentric distances (4 kpc < R$_{\rm GC}$ < 16 kpc). 
The literature ages, obtained from isochrone fitting of the full cluster sequence, of the clusters located at R$_{\rm GC}$ > 7 kpc on average agree well with the ages derived from the relations between abundance ratios and ages. 
However, the ages derived for the innermost open clusters, R$_{\rm GC} <$ 7 kpc, were overestimated with respect to those from the literature. In other words, at a given age, clusters located in the inner disc had a lower (or equal) Y abundance than clusters at larger Galactocentric radii, while their Mg abundance increased, producing a lower [Y/Mg]. 
C20 tentatively explained the behaviour of the inner disc stellar populations as due to a different star formation history (SFH) in different Galactocentric distances, combined with a non-monotonic metallicity dependence of the s-process stellar yields, and assuming a decreasing efficiency of the s-process at high metallicity. 
With a Galactic chemical evolution (GCE) model  \citep{magrini09} and a set of empirical yields in which the production of Y at high metallicity was depressed, C20 were able to reproduce the observed trends. 

We here provide a robust theoretical explanation of this decrease in the s-process efficiency with increasing metallicity, and how this can reconcile the observations with the results of a GCE model. 
In Section~\ref{sec:yields} we describe the assumptions under which new AGB models including magnetic mixing were built. In Section~\ref{sec:GCE} we compare the results of a GCE with the abundances and ages of a sample of open clusters observed in the Gaia-ESO survey that were presented in C20 and \citet{magrini18}. 
In Section~\ref{sec:conlusion} we summarise our results and conclude.

\section{Yields from magnetic AGB stars}
\label{sec:yields}
Stellar yields from asymptotic giant branch (AGB) stars play a major role in constructing a GCE model. In a previous work (C20), AGB yields from different authors were adopted, including those available in the FRUITY database\footnote{http://fruity.oa-abruzzo.inaf.it/} \citep{cris11,cris15}. FRUITY models, coupled to rotating massive star models from \citet{cl2018}, provide a good fit to the distribution of s-process elements in the solar spectrum \citep{pra2020}. However, extremely precise ($\approx$ 10\%) isotopic ratio measurements in presolar SiC grains \citep{liu18} demonstrated that the neutron density of FRUITY models is likely overestimated. In low-mass AGB stars, neutrons are mainly provided by a $^{13}$C-rich thin layer, the so-called $^{13}$C pocket, which forms at the bottom of the convective envelope after each third dredge-up episode (see e.g. \citealt{cris09}). The extension of the pocket and its $^{13}$C enrichment depend on the algorithm that is adopted to handle the convective or radiative interface. FRUITY models are calculated by applying a simple exponentially decreasing profile of the convective velocities at the inner border of the convective envelope \citep{cris09}. Recently, these models have been revised by considering the mixing triggered by magnetic fields \citep{vescovi20}. This led to the formation of a more extended and flatter $^{13}$C pocket with a lower $^{13}$C concentration. We refer to \citet{vescovi20} for the details of this implementation. New magnetic models better reproduce presolar grain experimental data by matching almost all the isotopic ratios. \citet{vescovi20} presented three models with the same initial mass ($M = 2~M_\odot$) and different metallicities ($Z=10^{-2}$, $Z = Z_\odot$ , and $Z = 2\times 10^{-2}$). To better identify the yttrium trend as a function of the metallicity, we extended the $2~M_\odot$ set by calculating three additional models ($Z = 3 \times 10^{-3}$, $Z = 6 \times 10^{-3}$ , and $Z = 3\times 10^{-2}$). As a reference term, we also calculated the $Z = 3\times 10^{-2}$ FRUITY model, which is not currently included in the database.

\section{Time evolution of the yttrium abundance: effect of magnetic mixing at high metallicity}
\label{sec:GCE}

We included the new set of Y stellar yields in the GCE described in \citet{magrini09} and \citet{maiorca12}, which is a multi-phase multi-zone semianalytic chemical evolution model based on the work of \citet{feerrinigalli88}, which has been applied by \citet{travaglio99}, \citet{molla05}, \citet{magrini07}, and \citet{molla15,molla16,molla19}, for example. 
For the production of the first- and second-peak neutron capture elements in massive stars, we considered the yields of \citet{cl2018}. 

\begin{figure}
   \centering
   \includegraphics[width=0.50\textwidth]{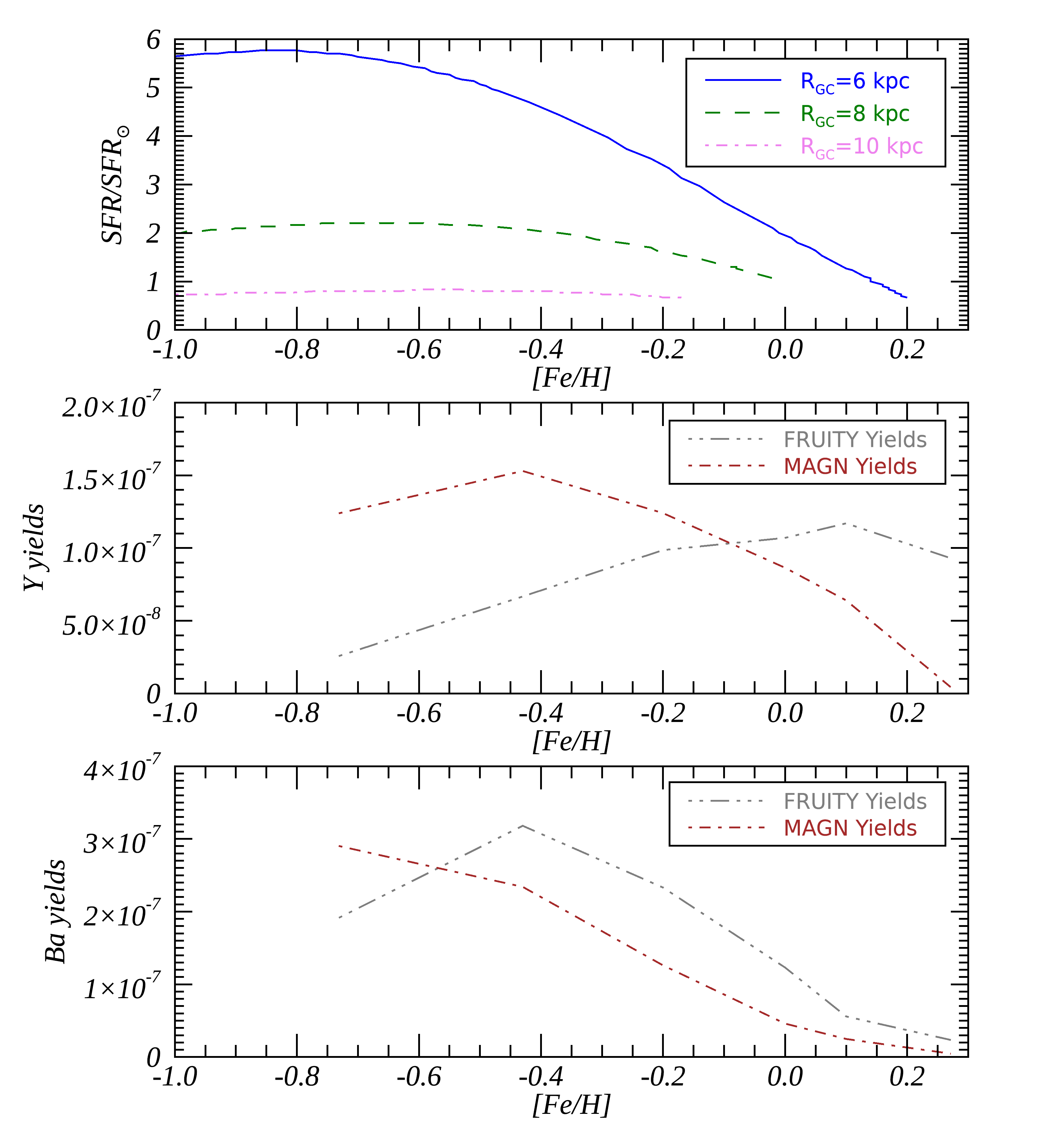}
   \caption{SFR and s-process yields as a function of metallicity. Upper  panel: SFR normalised to SFR at solar R~$_{\rm GC}$ at the present time computed with the GCE in the thin disc at R~$_{\rm GC}$ = 6, 8, 10~kpc. Middle panel: Net yttrium yields as a function of  [Fe/H] for a $2~M_{\odot}$ AGB star. Bottom panel: Net barium yield as a function of  [Fe/H] for a $2~M_{\odot}$ AGB star (grey: FRUITY yields; red: new MAGN yields).}  
\label{FigYields}%
    \end{figure}

In the upper panel of Fig.~\ref{FigYields} we show the star formation rate (SFR) normalised to the SFR at solar R$_{\rm GC}$ at the present time (SFR$_{\odot}$), predicted by our GCE at three Galactocentric radii (6-8 and 10~kpc).
As a consequence of the higher infall rate and the more efficient star formation in the inner parts of the disc, the SFR/SFR$_{\odot}$ is considerably higher at R$_{\rm GC}$ = 6~kpc than at larger radii. In addition, super-solar [Fe/H] are reached in the inner regions, unlike in the outermost regions, where lower values are obtained. 
In  the middle panel of Fig.~\ref{FigYields} we show the yttrium net yields of a $2~M_{\odot}$ as a function of [Fe/H] extracted from the FRUITY database and obtained with the new set of AGB models, in which the mixing triggered by magnetic fields is considered (hereafter MAGN). 
The two sets of yields differ strongly. This is caused by their different neutron-to-seed ratio. This ratio depends on the availability of free neutrons (numerator) and on the abundance of iron seed from which the s-process path starts (denominator). While the first quantity is of primary origin, the latter depends on the initial metallicity. 
In the bottom panel of Fig.~\ref{FigYields} we show the barium net of a $2~M_{\odot}$ from the  FRUITY database and with the new MAGN approximation. 
In general, MAGN models are characterised by a lower neutron-to-seed ratio because fewer neutrons are available to the synthesis of heavy elements. This has many important consequences {\it i)} at low metallicities, more yttrium is produced in MAGN models because in FRUITY models the s-process flow is strong enough to saturate the first s-process peak (Sr-Y-Zr region) and to move to the second peak (Ba-La-Ce region); {\it ii)} for higher [Fe/H], the two sets have opposite trends: MAGN models show an almost monotonically decreasing curve;  on the other hand, FURITY models first increase, reach a maximum at slightly super-solar metallicity, and finally start to decrease; {\it)} from low [Fe/H] to high [Fe/H], the net yttrium production of MAGN models decreases by roughly a factor of 30. In contrast, Y increases by about a factor of 4 in FRUITY models. The effect is less pronounced for the Ba yields at high [Fe/H] in the two sets of yields. 

\begin{figure}
   \centering
   \includegraphics[width=0.40\textwidth]{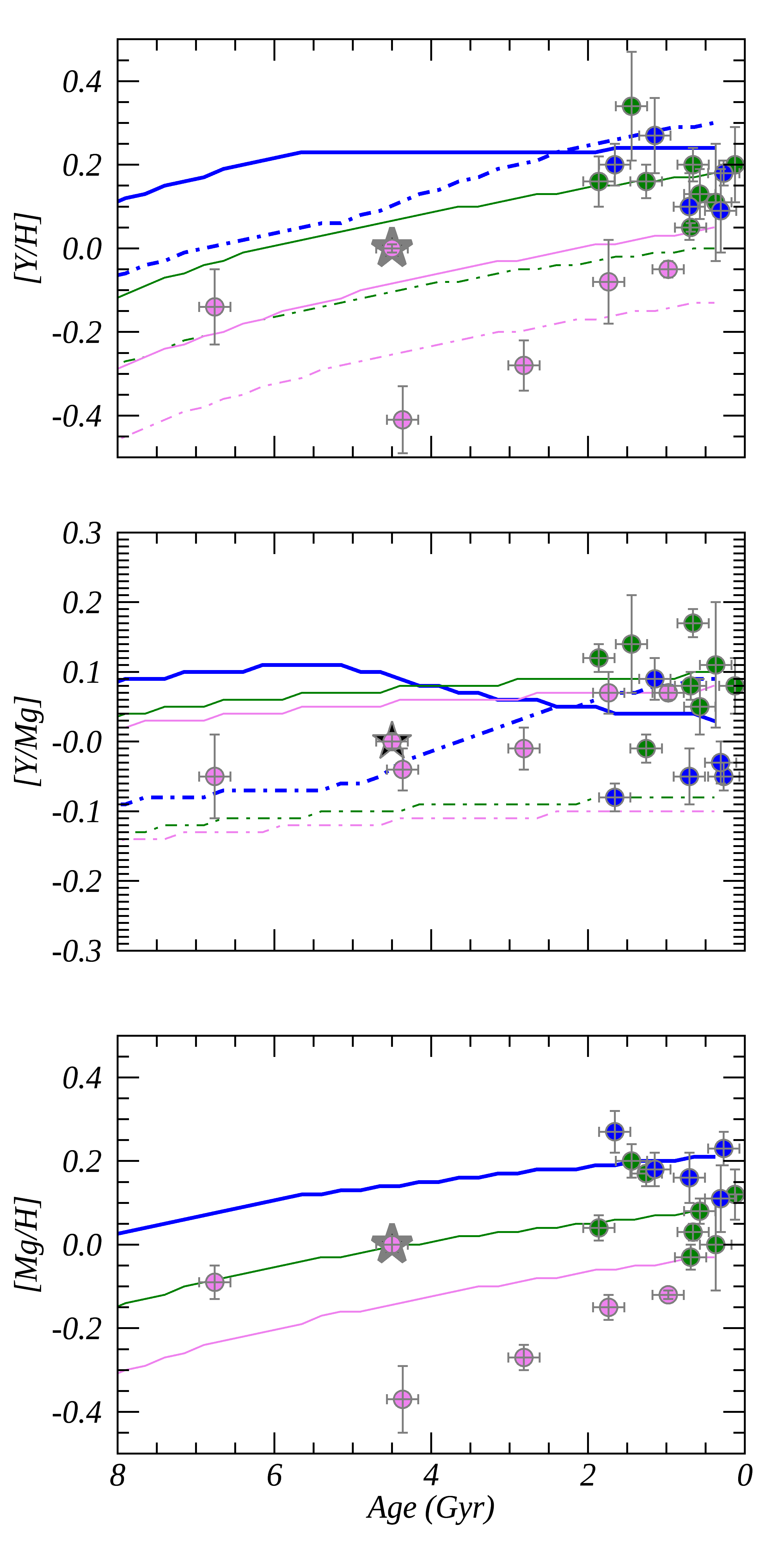}
   \caption{[Y/H], [Y/Mg], and [Mg/H] vs. age: Sample of Gaia-ESO {\sc idr5} clusters in three radial bins (R$_{\rm GC} <$ 6.5~kpc in blue, 6.5~kpc $<$ R$_{\rm GC}<$ 9~kpc in green, and R$_{\rm GC} >$ 9~kpc in pink) compared with results of the GCE for the thin disc at three R$_{\rm GC}$ = 6, 8, and 10~kpc with the MAGN yields (continuous curves) and with the FRUITY yields (dot-dashed lines). The ages and R$_{\rm GC}$ of open clusters are from \citet{cantat20}. The star marks the abundance ratio at the solar age and Galactocentric distance.}  
\label{FigGCE}%
    \end{figure}

It is therefore straightforward to hypothesise that the adoption of MAGN yields in the GCE has an important role in the inner disc, where super-solar metallicity is reached. 
Still, we caution that the currently available MAGN models only have initial masses $M=2~M_\odot$. In this work, AGB models with other stellar masses were scaled by conserving the original FRUITY dependence on the initial mass. This approximation has to be taken with caution because we did not yet explore the dependence of the mixing triggered by magnetic fields on the initial stellar mass (Cristallo et al., in preparation).\\ 
We plot the results in Fig.~\ref{FigGCE}, where we show the trend of [Y/H] versus age (upper panel)  by comparing our abundances and the ages of the open cluster sample, and the results of the models at R$_{\rm GC}$ = 6, 8, and 10~kpc. The open clusters are divided into three radial bins 
to be compared with the corresponding curves of the model. 
The observational data show us that in the innermost clusters yttrium no longer increases,
but that [Y/H] is comparable with the ratio in
solar neighbourhood clusters.   
We also show the theoretical curves obtained using the new MAGN models and with the FRUITY yields.
With the choice of the MAGN yields for 
Y, the GCE predicts a flatter curve in
the inner disc than in the outermost regions  for [Y/H] versus age, while with the FRUITY yields there would be further growth of [Y/H]. In the bottom panel we show the time evolution of [Mg/H], for which we
adopted the yields of \citet{cl04} for Z solar and Z = 0.006, extrapolated to 100 M$_{\odot}$. To be consistent with C20, we kept the same choice for the Mg yields to highlight
variations that are only due to the yields of the s-process elements.
As noted by \citet{prantzos18}, most literature Mg yields
underestimate the production of Mg, in particular those that include
rotating massive stars \citep[see also][]{romano10, magrini14}, and we therefore added an offset to the model curves
to reproduce the solar value. Unlike [Y/H], [Mg/H] continues
to grow in the inner part of the disc, both from the observational
side and from the results of the GCE. In the middle panel, we
show the time evolution of [Y/Mg]. The combination of an
increasing trend for Mg and a slightly decreasing trend for Y
decreases their ratio for the inner disc clusters with respect
to the clusters in the solar neighbourhood at a given age.
The results of the GCE model with the MAGN yields qualitatively reproduce the observed trend, predicting a lower [Y/Mg] than at R$_{\rm GC} \geq$ 8~kpc in the last 4~Gyr at R$_{\rm GC}$ = 6~kpc. On the other hand, the results of the GCE with standard FRUITY yields would produce a net increase of [Y/H] in the inner regions, which combined with the trend of [Mg/H], implies a higher [Y/Mg] in the inner regions with respect to the solar and outer regions.

\begin{figure}
   \centering
   \includegraphics[width=0.40\textwidth]{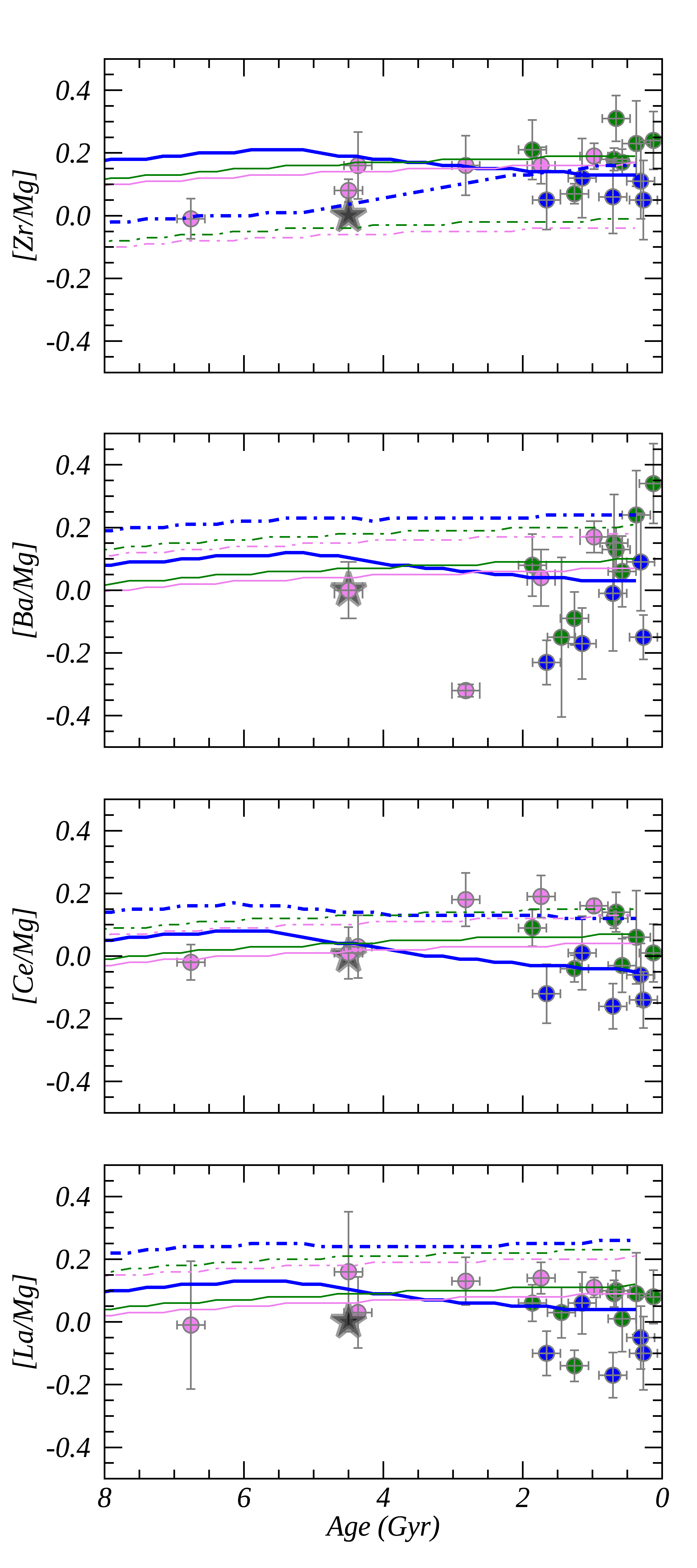}
   \caption{[Zr/Mg], [Ba/Mg], [Ce/Mg], and [La/Mg] vs. age. Symbols and colours as in Fig.~\ref{FigGCE}.  }  
\label{fig:allel}%
    \end{figure}

The results shown in Fig.~\ref{FigGCE} are similar to those obtained by C20 with a set of yields from \citet{maiorca12}, empirically rescaled to match the observations. In particular, C20 reduced the yields at super-solar metallicity by a factor 10. 
Now, the need for lower yields at high metallicity is instead based on a solid physical basis and justifies the previous empirical results. 
For the sake of completeness, in Fig.~\ref{fig:allel} we report the time evolution of  the other  s-process elements versus Mg observed in the open clusters of the {\sc idr5} of the Gaia-ESO survey, with data available in \citet{magrini18}. The five elements (including Y, shown in Fig.~\ref{FigGCE}), Zr and Y (belonging to the first s-process peak), and Ba-La-Ce (second s-process peak), show a similar behaviour: inner and solar Galactocentric clusters reach a similar enrichment in [El/Mg] despite the much more intense star formation in the inner part of the disc. 
We compare the observations with the results of the GCE models in which, as for Y, we include the MAGN yields. For comparison, we also include the results of the GCE with the FRUITY yields. The behaviour of Zr is very similar to that of Y because they both belong to the first peak. 
For elements belonging to the second s-process peak, we also find that inner disc clusters reach the same enrichment as their solar counterparts. GCE curves computed with MAGN models reproduce the observed trends well, further confirming the robustness of the new adopted physical prescription for magnetic mixing in AGB stars. Another striking achievement is the adequate reproduction of the solar abundances at the epoch and the birith place of the Sun ($\sim$8~kpc and $\sim$4.5~Gyr ago; marked by a star in the plot) for all the studied elements. These elements would be slightly overestimated for the second-peak elements when adopting FRUITY yields (see e.g. \citealt{pra2020}). The difference between the two sets arises from the fact that in the metallicity range we explored, MAGN models also produce fewer elements belonging to the second $s$-process peak than FRUITY models (see the lower panel of Fig.~\ref{FigYields}). 
 Further exploration of the low-metallicity regime is needed, as is the dependence on the magnetic mixing as a function of the initial mass (Vescovi et al. in preparation; Cristallo et al. in preparation). Moreover, we also plan to evaluate the possible effects caused by stellar migration (see e.g. \citealt{minchev19} and \citealt{cri2020}).

Data on additional open clusters from the last Gaia-ESO data release, covering wider ranges in ages and distances, will be available in Spring 2021. They will allow us to complete the comparison between models and observations by including  more s-process and r-process elements (Van der Swaelmen et al. in preparation; Viscasillas Vazquez et al. in preparation), which will for the first time provide spatially resolved relations of age and chemical clocks in the Galactic disc.

\section{Conclusions}

We propose a theoretical explanation of the behaviour of the slow neutron capture elements, in particular Y, in the inner part of the Galactic disc. 
In a recent paper, C20 confirmed the existence of a tight relation of age and [Y/Mg] in the solar neighbourhood. The failure to apply this relationship to the open cluster population in the innermost part of the Galactic disc has raised some questions about the efficiency of the s-process at high metallicity. C20 found an empirical solution that in a GCE model introduced reduced yields for the production of Y from AGB stars at high metallicity. 

Recently, \citet{vescovi20} proposed a new paradigm for mixing induced by magnetic buoyancy in AGB stars. With respect to this work, we expanded the metallicity range of our theoretical investigation and highlighted a new relationship between metallicity and high-mass element yields. With a focus on yttrium, we applied these new AGB yields in a GCE model and were able to reproduce the observed trend at different Galactocentric distances. 
In particular, we can qualitatively explain the change in slope and intercept in the relations of age and [Y/Mg]. 
We confirm that no single relationship of age and [Y/Mg] is valid in the whole Galaxy, but that it varies according to the SFR and the metallicity.  For limited regions, in particular the solar neighbourhood, the relations of age and abundance ratios built with solar twins are still valuable tools for measuring stellar ages. We aim at extending these relations including a further variable that is the birth radius in the disc (Viscasillas Vazquez et al. in preparation).

\label{sec:conlusion}
 
\begin{acknowledgements}
We thank the referee, Thomas Masseron, for his constructive report than has allowed us to improve the quality of the paper.  
LM, GC, MVdS acknowledge the funding from MIUR Premiale 2016: MITiC. MVdS and LM thanks the WEAVE-Italia consortium. LM acknowledge the funding from the INAF PRIN-SKA 2017 program 1.05.01.88.04. CVV and LM thank the COST Action CA18104: MW-Gaia. LS acknowledges financial support from the Australian Research Council (discovery Project 170100521) and from the Australian Research Council Centre of Excellence for All Sky Astrophysics in 3 Dimensions (ASTRO 3D), through project number CE170100013.  
DV acknowledges the financial support from the German-Israeli Foundation (GIF No. I-1500-303.7/2019).
This research has made use of NASA’s Astrophysics Data System Bibliographic Services.  
\end{acknowledgements}

%
%
\bibliographystyle{aa}
\bibliography{Bibliography}

\end{document}